\documentclass{iopart}
\usepackage{epsfig}
\begin{document}
\title{Double-impulse magnetic focusing of launched cold atoms}
\author{Aidan S Arnold\dag, Matthew J Pritchard\ddag, David A Smith\ddag~and Ifan G Hughes\ddag}
\address{\dag~Department of Physics, University of Strathclyde, Glasgow, G4 0NG, UK}
\address{\ddag~Department of Physics, Rochester Building, University of
Durham, South Road, Durham, DH1 3LE, UK}
\ead{i.g.hughes@durham.ac.uk}
\date{\today}
\begin{abstract}
We have theoretically investigated 3D focusing of a launched cloud
of cold atoms using a pair of magnetic lens pulses (the
alternate-gradient method). Individual lenses focus radially and
defocus axially or vice-versa. The performance of the two possible
pulse sequences are compared and found to be ideal for loading both
`pancake' and `sausage' shaped magnetic/optical microtraps. It is
shown that focusing aberrations are considerably smaller for
double-impulse magnetic lenses compared to single-impulse magnetic
lenses. An analysis of the clouds focused by double-impulse
technique is presented.
\end{abstract}

\section{Introduction}
The field of atom optics~\cite{Adams94} has undergone a dramatic
expansion in the last two decades, largely as a consequence of the
development of laser-cooling techniques~\cite{Adams97}, and the
routine production of atoms at microKelvin
temperatures~\cite{nobel97}. Paramagnetic cold atoms can be
manipulated  with the Stern-Gerlach force~\cite{Hinds}. To date, the
Stern-Gerlach force has been used to realise a variety of atomic
mirrors for both cold \cite{cold} and Bose condensed
atoms~\cite{BECmirror}. This paper, however, concentrates on the
formation of magnetic lenses for cold atoms. In comparison to a
ballistically expanding unfocused cloud, a magnetically focused
cloud can lead to density increases (or conversely temperature
decreases) of many orders of magnitude. Applications include: atom
lithography~\cite{meschede}; transferring cold atoms from a Magneto
Optical Trap (MOT) to a remote vacuum chamber of lower background
pressure~\cite{Szymaniek99}; cold low-density atomic sources for
fountain clocks \cite{clocks}.

The first demonstration of 3D focusing with pulsed magnetic lenses
was conducted by Cornell~{\it et al.}~\cite{mon1}. The group of
Gorceix have made experimental and theoretical studies of cold atom
imaging by means of pulsed magnetic fields~\cite{Marec, Gor}.
However, neither work addressed the optimum strategy for achieving a
compact focused cloud, nor the limiting features for the quality of
their atom-optical elements.

Recently we provided a theoretical analysis of 3D focusing of
weak-field-seeking cold atoms using a single magnetic
pulse~\cite{matt1}. Lens designs for 1D and 3D were presented that
minimise aberrations due to the lens potential's departure from the
perfect parabolic case. Single-impulse 3D focusing has been
experimentally achieved at Durham University, the results of which
can be seen in a forthcoming publication~\cite{Smith05}.

The scope of this paper is to investigate theoretically and numerically the limiting
factors to the quality and size of the final image obtained in double-impulse magnetic
focusing experiments; to identify the sources of aberration; and to discuss schemes for
minimising their effect. We will show that both single- and double-impulse lenses yield a
magnetically focused cloud with a bimodal distribution consisting of a highly diffuse
outer cloud, as well as a core cloud which can be orders of magnitude denser than the
initial atomic sample. This core cloud is therefore ideal for remotely loading tight
traps with relatively small depth, e.g. miniature magnetic guides~\cite{minmagguide},
atom chips~\cite{Hinds,atomchips} and optical dipole traps \cite{dipole}.

The remainder of the paper is organised as follows: Section 2
outlines the theory of how to achieve the desired magnetic fields;
Section 3 contains an analysis of magnetic imaging and minimising
the final cloud size; Section 4 describes and contrasts the spatial
performance of different magnetic lenses; Section 5 considers
experimentally relevant parameters for alternate gradient lenses;
Section 6 contains a discussion and concluding remarks.

\section{Alternate-gradient lens theory}

An atom in a magnetic field of magnitude $B$ experiences a magnetic
dipole interaction energy  of
 $U=-\mu_{\zeta}B$, where $\mu_{\zeta}$ is the projection of the atom's magnetic moment onto the field direction.
Provided that Majorana spin-flip transitions \cite{petr} occuring in
field zeros are avoided and the rate of change of direction of the
field is less than the Larmor frequency the magnetic moment
adiabatically follows the field.

The Stern-Gerlach force is $\vec{F}_{\rm SG}=-\nabla U=\nabla
(\mu_{\zeta}B)$.  The ensemble (of alkali metal atoms) can be
optically pumped into either a strong-field-seeking state with
$\mu_{\zeta} = \mu_{\rm B}$ (where $\mu_{\rm B}$ is the Bohr
magneton), or into a weak-field-seeking state with $\mu_{\zeta} =
-\mu_{\rm B}$. In low field, where the quantum numbers $F$ and
$m_{F}$ are good, these states are the stretched states $\vert
F=I+1/2, m_{F}=\pm F \rangle$. Atoms in these states have a magnetic
moment which is independent of field, consequently the Stern-Gerlach
force takes the simpler form $\vec{F}_{\rm SG}=\pm\mu_{\rm B}\nabla
B$ --- i.e. the focusing of the atoms is governed by the gradient of
the magnetic field magnitude only.

The choice of whether atoms in weak or strong-field seeking states are launched depends on the
particular application. We discussed extensively in~\cite{matt1} the focusing of atoms in
weak-field-seeking states. This was because single-impulse 3D imaging of strong-field-seeking
states requires a maximum of the magnetic field in free space, which is forbidden by Earnshaw's
theorem \cite{earnshaw}.

In this paper magnetic lenses centred on the point $\{0,0,z_{\rm c}\}$ are considered, with a second order magnetic field
magnitude of the form:
\begin{equation}
B(x,y,z)=B_{0}+\frac{B_2}{2}\left(-x^2/2-y^2/2+(z-z_{\rm
c})^2\right). \label{1D}
\end{equation}
$B_{0}$ and $B_{2}$ are the bias field and the field curvature,
respectively. Substituting this into the Stern-Gerlach force
expression results in an atom of mass $m$ experiencing a harmonic
acceleration about $\{0,0,z_{\rm c}\}:$
\begin{equation}
\textbf{a}= -{\omega}^2\{-x/2,-y/2,(z-z_{c})\},
 \label{force1D}
\end{equation}
where ${\omega}^2=\mu_{\zeta}B_{2}/m$ is a measure of the power of
the lens. The axial curvature is twice the magnitude of, and
opposite in sign to, the radial curvature, ${\omega_\mathrm{z}}^2=-2
{\omega_\mathrm{r}}^2$. Note that lens curvature in all three
spatial dimensions is reversed if the sign of either $\mu_{\zeta}$
or $B_2$ is reversed. For simplicity from this point on, only the
case $\mu_{\zeta}=-\mu_B$ i.e. weak-field-seeking atoms will be used
and lens curvature is modified solely via $B_2$.

\begin{figure}[ht]
\begin{center}
\vspace{-5mm} \epsfxsize=\columnwidth \epsfbox{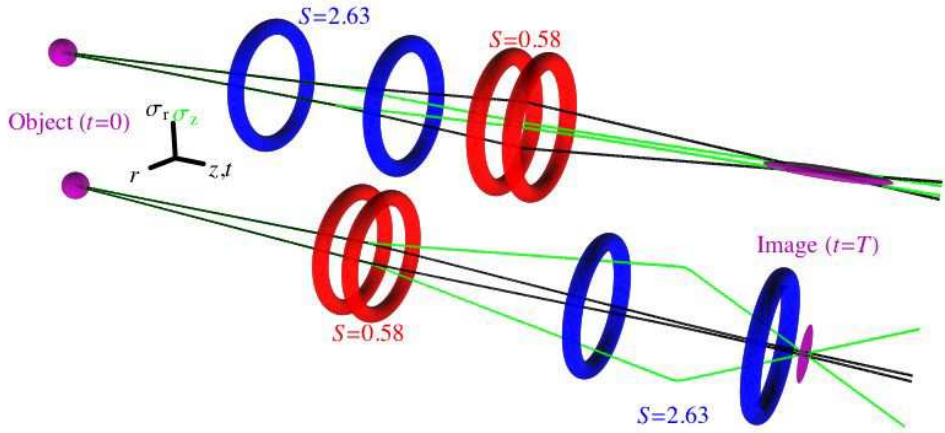}
\vspace{-13mm} \caption{\label{schematic} The principle of alternate
gradient focusing. The upper image, Strategy AR, shows the evolution
of the axial $(\sigma_z,$ green) and radial $(\sigma_r,$ black)
cloud radii when an axially converging (radially diverging) lens
precedes an axially diverging (radially converging) lens and leads
to a sausage-shaped cloud $(\xi>1)$. In the lower image, Strategy
RA, the lens order is reversed, leading to a pancake-shaped cloud
$(\xi<1)$. The effects of gravity (in the $-z$ direction along the
coil axis) are not shown, but are included in simulations. Due to
the time reversal symmetry of optics, the lens system also works
backwards.}
\end{center}
\end{figure}

We refer to a lens with a field expansion of the form $B$ as a
one-dimensional lens, as it can be used either to focus axially
(with simultaneous radial defocusing), when $B_2$ is positive, or to
focus radially (with simultaneous axial defocusing) when $B_2$ is
negative. In order to achieve a 3D focus with the lenses of
equation~(\ref{force1D}), an axially converging lens pulse must be
followed by an appropriately timed axially diverging lens (or vice
versa). This is referred to as the ``alternate gradient" focusing
method, and has the advantage of being able to focus both weak-field
and strong-field seeking atoms. This method is used extensively in
particle accelerators~\cite{altgrad}, for focusing polar
molecules~\cite{meijer}, and is shown schematically in
Figure~\ref{schematic}. Useful parameters for describing the
evolution of a Gaussian atomic cloud are the axial $(\sigma_z)$ and
radial $(\sigma_r)$ rms cloud radii (the standard deviations), as
well as their aspect ratio $\xi=\sigma_z/\sigma_r.$

 As shown in reference~\cite{matt1} there exist optimal configurations for
realising radial and axial focusing lenses. These are achieved with
a pair of separated coaxial coils, where both coils have equal
current with the same sense. An important lens parameter is the
relative separation $S$ of the coils in units of the coil radius.
The harmonicity of a radially-focusing lens is optimized if $S=0.58$
(red lenses in Figure~\ref{schematic}); whereas the harmonicity of
an axially-focusing lens is optimized if $S=2.63$ (blue lenses in
Figure~\ref{schematic}). In the remainder of this work it is assumed
that these optimized lenses are used.

\section{Magnetic impulses and the ABCD formalism}\label{imagesec}
The separable equations of motion for an axially and cylindrically
symmetric coil system lead to a lens that is harmonic in 3D with an
acceleration given by equation  (\ref{force1D}), which allows the
motion in each cartesian dimension to be treated as a separate
simple harmonic equation. The influence of a magnetic lens can be
described by $\mathcal{ABCD}$ matrices, as outlined in
\cite{Gor,matt1}. The initial and final position and velocity of an
atom along a given Cartesian axis, say $x$, are related via the
equation:
\begin{equation}
\left( \begin{array}{c} x_{\rm f}  \\
          v_{x_{\rm f}} \end{array} \right)=\left( \begin{array}{cc}
\mathcal{A} & \mathcal{B}  \\
\mathcal{C}  & \mathcal{D}
\end{array} \right)
\left( \begin{array}{c} x_{\rm i}  \\
          v_{x_{\rm i}} \end{array} \right).
\label{eq:ABCD}
\end{equation}

A `thick' converging lens of strength $\omega$ (with ${\rm
Im}(\omega)=0)$ and physical duration $\tau$ is actually equivalent
to the `thin' lens $\mathcal{ABCD}$ transformation:
\begin{equation}
\!\!\!\!\!\!\!\!\!\!\!\!\! \left(
\begin{array}{cc}
\cos(\omega \tau) & \frac{1}{\omega}\sin(\omega \tau) \\
-\omega\sin(\omega \tau) & \cos(\omega \tau)
\end{array} \right)=
\left( \begin{array}{cc}
1 & \tau'/2 \\
0 & 1
\end{array} \right)
\left( \begin{array}{cc}
1 & 0 \\
\mathcal{C} & 1
\end{array} \right)
\left( \begin{array}{cc}
1 & \tau'/2 \\
0 & 1
\end{array} \right),
\label{thincon}
\end{equation}
where $\mathcal{C}(\omega,\tau)=-\omega \sin(\omega \tau),$ and is
pre- and post-multiply by a translation matrix of half the effective
pulse width $\tau'(\omega,\tau)=\frac{2}{\omega}\tan(\frac{\omega
\tau}{2}).$ The notation of primes is used to denote times in the
`thin' lens representation. To model a diverging lens make the
transformation $\omega\rightarrow \pm {\rm i}\omega$ in equation
(\ref{thincon}) -- i.e.~$\mathcal{C}=\omega\sinh(\omega \tau)$ and
$\tau'=\frac{2}{\omega}\tanh\frac{\omega \tau}{2}.$ The `equivalent
time' of the lens $\tau'$ is not the same as the real experimental
pulse duration of $\tau$.

\subsection{Double impulse magnetic lenses - the parabolic case}
\begin{figure}[ht]
\begin{center}
\epsfxsize=\columnwidth \epsfbox{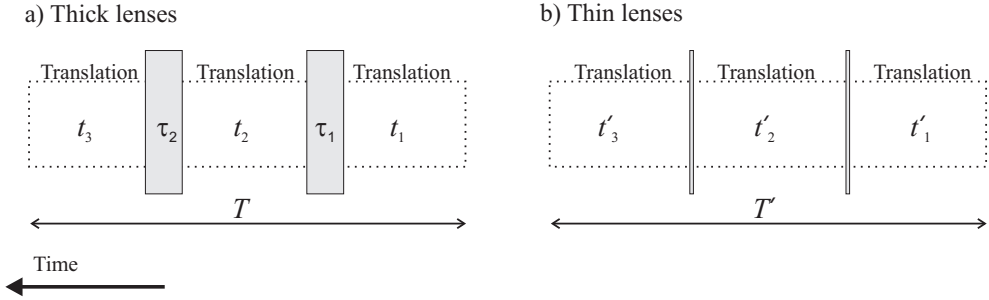}
 \caption{\label{timings} A schematic diagram showing the timing sequences for the two frames of reference.
 (a) shows the thick lens or lab frame with the time durations for each stage and (b) shows the mathematically
 equivalent thin lens representation that is used in the calculations. Note that the direction of time runs
 right to left so that it visually mirrors the system matrix layout in equation~(\ref{doublelens}).}
\end{center}
\end{figure}

A double lens system, see Figure~\ref{timings}, comprising lenses of
strength and duration $\omega_1,$ $\tau_1$ (starting after a time
$t_1$) and $\omega_2,$ $\tau_2$ (starting a time $t_2$ after the
first lens) is modelled by using the following $\mathcal{ABCD}$
matrix sequence:
\begin{equation}
\!\!\!\!\!\!\!\!\!\!\!\!\!\!\! \!\!\!\!\!\!\!\!\!\!\!\!\!\!\!
 \left( \begin{array}{cc}
\mathcal{A} & \mathcal{B}  \\
\mathcal{C}  & \mathcal{D}
\end{array} \right)=\left( \begin{array}{cc}
1 & t_3' \\
0 & 1
\end{array} \right)
\left( \begin{array}{cc}
1 & 0 \\
\mathcal{C}_2 & 1
\end{array} \right)
\left( \begin{array}{cc}
1 & t_2' \\
0 & 1
\end{array} \right)
\left( \begin{array}{cc}
1 & 0 \\
\mathcal{C}_1 & 1
\end{array} \right)
\left( \begin{array}{cc}
1 & t_1' \\
0 & 1
\end{array} \right),
\label{doublelens}
\end{equation}
i.e. a $t_1'=t_1+\frac{1}{2}\tau_1'$ translation, then a strength $\mathcal{C}_1$ thin
lens, a $t_2'=\frac{1}{2}\tau_1'+t_2+\frac{1}{2}\tau_2'$ translation, then a strength
$\mathcal{C}_2$ thin lens followed by a $t_3'=T'-t_1'-t_2'$ translation, where
$\mathcal{C}_j=-\omega_j\sin(\omega_j\tau_j)$ and
$\tau_j'=\frac{2}{\omega_j}\tan(\frac{\omega_j\tau_j}{2})$ for $j\in\{1,2\}$. The total
physical duration of the focusing, $T,$ is fixed, and the effective total time of the
double lens system is $T'=T-\tau_1-\tau_2+\tau_1'+\tau_2'$.

By multiplying the matrices of equation (\ref{doublelens}) together,
the final $\mathcal{ABCD}$ system matrix is obtained. An image (i.e.
a one-to-one map of position between the initial and final cloud) is
formed if the condition $\mathcal{B}=0$ is maintained. In this case
the spatial magnification $\mathcal{A}$ is the inverse of the
velocity magnification $\mathcal{D}$; a manifestation of Liouville's
theorem. The cloud extent along $x$ in a given plane is given by:
\begin{equation}
\sigma_{x_{f}}^{2} = (\mathcal{A}\sigma_{x_{i}})^2 +(\mathcal{B}
\sigma_{v_{x_{i}}})^2, \label{size}
\end{equation}
where $\sigma_{x_i}$ is the initial position standard deviation and
$\sigma_{v_{x_i}}$ is the initial velocity standard deviation. An
{\it image} is formed for the condition $\mathcal{B}=0$, but the
{\it smallest cloud size} occurs when one minimises the product of
the cloud extent for all 3 spatial dimensions
(i.e.~${\sigma_{r_{f}}}^{2} \sigma_{z_{f}}$). For single- and
double-impulse lens systems, the cloud size at the image plane and
the minimum cloud size do not correspond exactly, but they are
usually very similar. In the rest of the paper we will consider the
cloud size at the image plane $(\mathcal{B}=0)$, and thus
$\mathcal{A}$ corresponds to the magnification.

\subsection{Solving the matrix equations}
The important entries of the system matrix in equation
(\ref{doublelens}) are $\mathcal{A}$ and $\mathcal{B}$:
\begin{eqnarray}
\!\!\!\!\!\!\!\!\!\!\!\!\!\!\!\!\!\!\!\!\!\!\!\!\!\!\!\!\!\!\!\!\!\!\!\!\!\!\!\!\!\!
 \mathcal{A}=1 + \left(\mathcal{C}_1 +\mathcal{C}_2 \right) \left(
      T' - {t_1'} \right)  + \mathcal{C}_2 \left( -1 + \mathcal{C}_1 \left(
          T' - {t_1'} \right)  \right) {t_2'} - \mathcal{C}_1 \mathcal{C}_2 {{t_2'}}^2
          \\
\!\!\!\!\!\!\!\!\!\!\!\!\!\!\!\!\!\!\!\!\!\!\!\!\!\!\!\!\!\!\!\!\!\!\!\!\!\!\!\!\!\!
 \mathcal{B} = T' + \mathcal{C}_2\left( T' -{t_1'} - {t_2'} \right) \left( {t_1'} + {t_2'} \right)  +
  \mathcal{C}_1 {t_1'} \left( T' + \mathcal{C}_2\,\left( T' - {t_2'} \right) \,{t_2'} - {t_1'}\left( 1 +
   {\mathcal{C}_2}{t_2'} \right)
  \right),
\label{AB}
\end{eqnarray}
which are both second order in $t_1'$ and $t_2'$ (and hence also second order in $t_1$
and $t_2).$

To obtain an atom cloud which is focused in all 3 dimensions
requires that the first lens is axially converging (radially
diverging) and the second lens is axially diverging (radially
converging), or vice versa. Moreover, the radial (subscript $r$) and
axial (subscript $z$) spatial dimensions have different
$\mathcal{A}$ and $\mathcal{B}$ coefficients. If the two axial lens
strengths are $\omega_{1z}$ and $\omega_{2z},$ then equation
(\ref{force1D}) yields $\omega_{1r}={\rm i}\omega_{1z}/\sqrt{2}$ and
$\omega_{2r}={\rm i}\omega_{2z}/\sqrt{2}.$ A 3D image is formed when
equation~(\ref{AB}) is set equal to zero for both the radial and
axial directions.

In reference \cite{matt1} the density increase from a single-impulse
isotropic 3D harmonic lens, $\lambda^3/(1-\lambda)^3,$ was
characterised by $\lambda,$ the equivalent time of the thin lens,
${t_1}',$ relative to the total equivalent focus time $T'.$ Note
that for the anisotropic lenses in this paper the
\textit{equivalent} (i.e. thin lens) timing of a lens in the radial
and axial direction is different. For this reason we will
characterise alternate-gradient lensing with the parameters
$\{\lambda_1,\lambda_2\}=\{t_1+\tau_1/2,t_1+\tau_1+t_2+\tau_2/2\}/T,$
corresponding to the mean times of the first and second magnetic
impulses relative to the total experimental lensing time $T.$ We use
this labelling of $\{\lambda_1,\lambda_2\}$ if $\omega_{1r}$ is real
(the first lens is radially converging), and we swap the definitions
of $\lambda_1$ and $\lambda_2$ if $\omega_{1r}$ is imaginary (the
first lens is radially diverging).

Modelling an experiment being conducted at Durham University
\cite{Smith05}, we fix $T=212\,\rm{ms}$. The cold atom cloud has an
isotropic initial spatial and velocity distribution with 1D standard
deviations of $\sigma_{R}=0.4\,{\rm mm}$ and $\sigma_V=4.4\,{\rm
cm/s}$ (i.e. a temperature of~$20\,\mu{\rm K}$) respectively. The
coils are assumed to have a $4\,{\rm cm}$ radius with 10,000
Amp-turn current in each coil. The two lens combinations in
Figure~\ref{schematic} are shown in the table below with the
resulting angular frequencies.

\begin{table}[!th]
\begin{tabular}{|l|c|c|c|c|c|c|c|}
\hline
Strategy&1st lens&$\omega_{1r}$&$S_1$&2nd lens&$\omega_{2r}$&$S_2$&$\xi$\\
\hline

AR&Axial focus&$58{\rm
i}\,{\rm rad~s}^{-1}$&2.63&Radial focus&$97\,{\rm rad~s}^{-1}$&0.58&$>1$\\

RA&Radial focus&$97\,{\rm rad~s}^{-1}$&0.58&Axial focus&$58{\rm
i}\,{\rm rad~s}^{-1}$&2.63&$<1$\\

\hline
\end{tabular}
\caption{The two different alternate-gradient strategies modelled.
The $\omega$'s are the lens strengths, $S$'s are the coil
separations and $\xi=\sigma_z/\sigma_r$ is the cloud aspect ratio.}
\end{table}

\begin{figure}[ht]
\begin{center}
\epsfxsize=\columnwidth \epsfbox{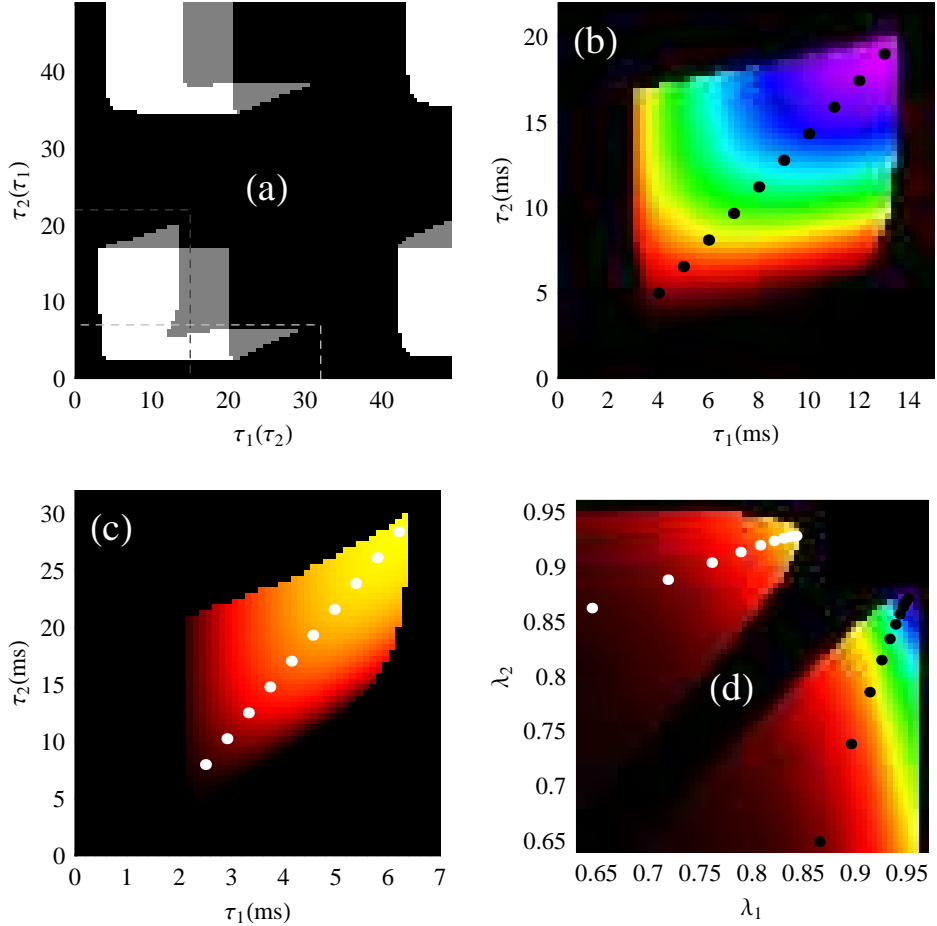}
 \caption{\label{sols} Image (a) shows the number of solution pairs (black=0, grey=1,
white=2) for $(t_1,t_2)$ as a function of $\tau_1$ and $\tau_2$
($\tau_2$ and $\tau_1$) in ms for Strategy AR (Strategy RA). The two
dashed regions of the `solution island' lead to the highest relative
density increases, shown in (b) and (c). The relative density
increase (equation (\ref{rhoinc})) if one images a cloud of atoms
using: (b) Strategy AR, (c) Strategy RA. The maximum relative
density increases are 1100 (320), for a $\xi=17$ sausage
($\xi=0.094$ pancake) shaped cloud, for images (b) and (c)
respectively. The results of (b) and (c) are combined in (d), the
relative density increase in terms of $\lambda_1$ and $\lambda_2$
(the mean times of the radially converging and radially diverging
impulses relative to $T$). The points in images (b)-(d) are used
later as a sample in simulations.}
\end{center}
\end{figure}

For a range of values of $\tau_1,$ and $\tau_2,$ we then solve the
radial and axial simultaneous equations (\ref{AB}) (i.e.
$\mathcal{B}_r=0,$ $\mathcal{B}_z=0$) to determine $t_1$ and $t_2.$
Although both $\mathcal{B}_z$ and $\mathcal{B}_r$ are quadratic in
$t_1$ and $t_2,$ substitution for either of these variables leads to
a final sextic polynomial equation. This must therefore be solved
numerically and leads to six solution pairs $(t_1,t_2).$ Only
solution pairs with real times $0\leq t_1,t_2 \leq T$ satisfying the
condition $t_1+\tau_1+t_2+\tau_2\leq T$ are considered. The number
of $(t_1,t_2)$ solution pairs as a function of $\tau_1$ and $\tau_2$
is shown in Figure~\ref{sols}(a). These $(t_1,t_2)$ solution pairs
can then be used to calculate the relative increase in atomic
density of a cold atom cloud. From equation~(\ref{size}) the
relative density increase of the image is thus:

\begin{equation}
\rho_{3D}=\frac{{\sigma_R}^3}{((\mathcal{A}_r \sigma_{R})^2
+(\mathcal{B}_r \sigma_{V})^2)\sqrt{(\mathcal{A}_z\sigma_{R})^2
+(\mathcal{B}_z \sigma_{V})^2}}\;\rightarrow
\frac{1}{{\mathcal{A}_r}^2\mathcal{A}_z}.
\label{rhoinc}\end{equation} Where the arrow indicates the limit
$\mathcal{B}=0$. The dimensionless relative density increases
obtained for both strategies are shown in Figure~\ref{sols}(b,c).
These plots are then effectively combined in Figure~\ref{sols}(d) by
inverting $\tau_1,\tau_2$ to find the relative density increase as a
function of the parameters $\lambda_1,\lambda_2$ which are the mean
relative times of the radially diverging and converging lens.

\section{A measure of the quality of the focus}

The attributes of parabolic lenses are unimportant, unless it can be
shown that experimentally realistic lenses are sufficiently
parabolic for such an approximation to be appropriate. At the end of
\cite{matt1} there appeared to be a major difference between the
parabolic approximation and real lenses. To some extent, this may
have been due to the way the lens properties were measured.

In \cite{matt1} numerical integration of the forces arising from the
magnetic fields due to current loops (generated by the Biot-Savart
law) was used, to track the trajectories of several $(\approx 1000)$
simulated atoms. The initial positions and velocities of the atoms
were randomly assigned, weighted according to isotropic Gaussian
spatial and velocity distributions with 1D standard deviations of
$\sigma_{R}=0.4\,{\rm mm}$ and $\sigma_V=4.4\,{\rm cm/s}$ (as
discussed in the previous section). The way in which the harmonicity
of a lens was measured was to compare the expected harmonic focus
size to the rms radii of the simulated atom cloud at the time of the
harmonic focus. The important drawback of this rms approach is that
the final location of atoms after a magnetic lens is highly
nonlinear with respect to initial conditions. An atom with a
velocity in the wings of the initial Gaussian distribution will
experience highly anharmonic lensing, as it will be far from the
centre of the lens during the magnetic impulse. Thus a few atoms can
completely alter the rms width of the cloud.

Another method to quantify the focus, adopted here, is to monitor
the fraction of the atoms entering the focus region of a purely
harmonic lens. The initial radial and axial cloud standard
deviations are $\sigma_R$, so the final standard deviations for a
harmonic lens are $\sigma_r=\mathcal{A}_r \sigma_R$ and
$\sigma_z=\mathcal{A}_z \sigma_R.$ By renormalising the dimensions
so that the radial and axial dimensions are measured in terms of
these final focus standard deviations, then a sphere with radius
$R_0=1.53817$ defined by
\begin{equation}
\frac{\int_{0}^{R_0}{r^2e^{-\frac{r^2}{2}}}dr}{\int_{0}^{\infty} {r^2e^{-\frac{r^2}{2}}}dr}=\frac{1}{2}
\end{equation}
will contain half of the atoms of the focused Gaussian distribution.
For numerical simulations the fraction of atoms entering this
harmonic focus is measured, and multiplied by twice the relative
density increase of a purely harmonic lens (i.e.
$2{\mathcal{A}_r}^{-2} {\mathcal{A}_z}^{-1}$) to get a measure of
the relative density increase afforded by a real lens.

Note that the centre of the harmonic focus region, as well as the
centre about which the rms radius is defined, is the final position
of an atom initially in the centre of the Gaussian position and
velocity distributions. This will lead to a slight underestimation
in the density increase regardless of which way it is defined (it
has been assumed the mean cloud position follows the initial
centre-of-mass). In addition the density can increase if the best
experimental 3D focus occurs at a time other than the best parabolic
lens focus time, but the focus time has been allowed to vary in the
simulations. For more details on the effects of gravity on launched
atoms, namely that even the centre-of-mass atoms will experience
time-varying radial and axial harmonic trap frequencies, see the
Appendix.

\subsection{Single-impulse focusing revisited}
In light of the above discussion, 3D single-impulse focusing
outlined in \cite{matt1} is briefly revisited, in particular
Strategy VI: the baseball coil system. This system consisted of a
square baseball coil with side lengths $w$ and a coil pair of radius
$w$ separated by $w$ that were coaxial with the baseball coil axis.
The current in the baseball coil was $10,000$~Amp-turns, and an
isotropic lens is formed when the coil pair has a current of
$1,541$~Amp-turns. In this paper we have made the value $w=4\,$cm,
leading to an angular frequency $\omega=62\,{\rm rad/s}$ in the
harmonic lens. This is to provide a better comparison between
single- and double-impulse techniques.

\begin{figure}[ht]
\begin{center}
\epsfxsize=\columnwidth \epsfbox{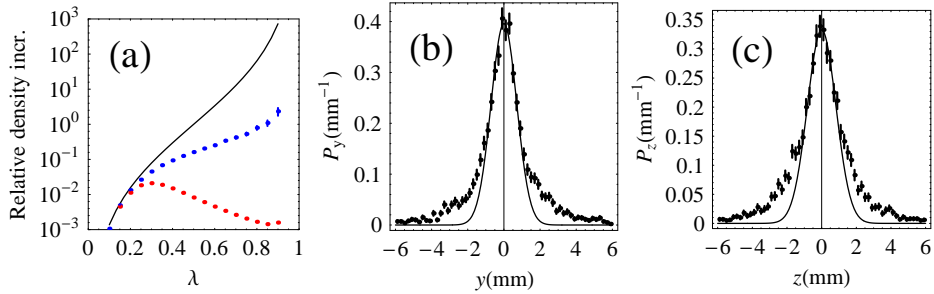}
 \caption{\label{baseball} Image (a) shows the relative density increase (with error bars, for a 5000 atom
simulation) based on: a purely harmonic lens (black curve), the
fraction of atoms in a real lens arriving at the harmonic focus
region (blue dots), the ratio of rms cloud volume before and after a
real lens (red dots).  Images (b) and (c) show the spatial
probability distributions at the focus (black dots) in the $y$ and
$z$ directions, respectively, where the non-Gaussian wings of the
distribution can be clearly seen. These two distributions are taken
from the $\lambda=0.5$ lens simulation, in which case $16\%$ of the
atoms are in the harmonic focus region. The Gaussian fits in (b) and
(c) (as well as the $x$ distribution) have an area of $\approx
70\%$.}
\end{center}
\end{figure}

In Figure~\ref{baseball} the relative density increase after the
baseball lens is plotted in terms of the parameter $\lambda$ (the
effective time of the baseball lens relative to $T$). The red dots
correspond to the relative density increase using rms widths for the
volume. The blue dots show the relative density increase as the
fraction of atoms in the harmonic focus zone times the harmonic
density increase. It is clear that we reach very different
conclusions based on whether the rms radius of the focused atomic
cloud, or the fraction of atoms which reach the harmonic focus are
considered.

By only looking at rms widths the optimum lens position occurs at
$\lambda=0.3,$ corresponding to a factor of 50 decrease in density.
However considering the fraction of atoms in the harmonic focus zone
times the harmonic density increase, the optimum position is now
$\lambda=0.9$. The relative density increase is 2.3, which
corresponds to 0.3$\%$ of the cloud focused to a density 729 times
greater than it was originally.

The source of this discrepancy between methods can be seen in
Figure~\ref{baseball} (b-c) where the non-Gaussian wings produce an
over estimate of the rms cloud width. The $y$ and $z$ standard
deviations for the Gaussian fits (black curves) are $730\,\mu$m and
$820\,\mu$m respectively; an almost isotropic distribution.

\subsection{Alternate gradient}

\begin{figure}[ht]
\begin{center}
\epsfxsize=\columnwidth \epsfbox{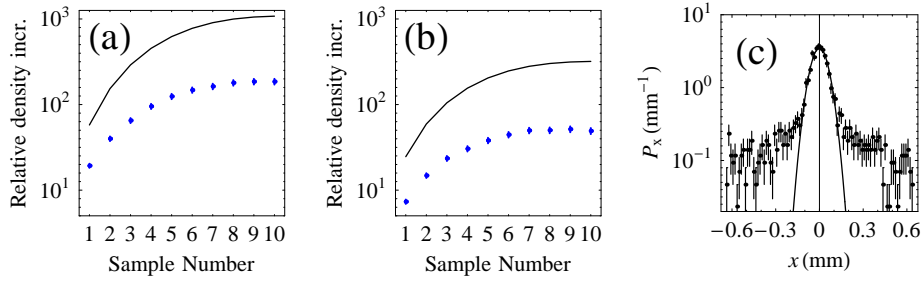}
 \caption{\label{altgrad} Images (a) and (b) use the $(\tau_1,\tau_2)$ co-ordinates illustrated in
Figure~\ref{sols}(b) and (c) to show the relative density increase
for alternate gradient lensing Strategies AR and RA, respectively.
There were 1000 atoms in the simulation and relative density
increases are shown for a pure harmonic lens (black curve), as well
as the relative density increase for the fraction of atoms in a real
lens arriving at the harmonic focus region (blue dots with error
bars). In image (c) the strong spatial bimodal nature of the $x$
focus for the leftmost point in (a) is clearly seen on a log scale.
The Gaussian fit (with $\sigma_x=56\,\mu$m) contains $49\%$ of the
3000 simulated atoms used in (c).}
\end{center}
\end{figure}

This section compares the alternate gradient numerical simulations
with the purely harmonic lenses of section~\ref{imagesec}. The
$(\tau_1,\tau_2)$ sample co-ordinates illustrated in
Figure~\ref{sols}(b-c) are used, in order to run numerical
simulations for the relative density increases illustrated in
Figure~\ref{altgrad}. We have not (cf. Figure~\ref{baseball}) used
the rms volume of the cloud to show the relative density increases
as these result in extremely low relative density increases
(typically $10^{-5}$ in (a) and $10^{-3}$ in (b)) that would reduce
the contrast in Figure ~\ref{altgrad}.

For both strategies the numerical simulations trace the shape of the
analytical relative density increase although aberrations result in
reduced increases. The maximum relative density increases in (a) and
(b) are 186 and 50 respectively. This is a marked improvement on the
single-impulse focusing, however the cloud distribution is no longer
isotropic. The harmonic focus aspect ratio has a range
$12\leq\xi\leq17$ in (a) and  $0.074\leq\xi\leq0.095$ in (b).

In certain applications, for example microtrap loading and
lithography, the sausage-shaped distribution with its reduced radial
spread could be beneficial. Figure \ref{altgrad} (c) plots the
distribution of a cloud focused via Strategy AR. The standard
deviations for the Gaussian fits to the core of the $x,$ $y,$ and
$z$ distributions are $56\,\mu$m, $56\,\mu$m and $850\,\mu$m; an
order of magnitude reduction in the radial direction compared with
single-impulse focusing.

Given comparable lens dimensions and strengths, double-impulse
magnetic focusing is far superior to single-impulse magnetic
focusing in terms of the relative density increases that can be
achieved by a fraction of the atoms. This result is in stark
contrast to the relative rms density increase of the entire cloud,
which would lead to the opposite conclusion. This spatial focusing
would find applications in lithography or sending the atomic cloud
through micro-sized apertures.

\section{Physical properties of clouds focused via the alternate gradient method}

For many experiments it is also important to consider changes to the
velocity distribution and hence the collision rate and phase-space
density. Unless the collision rate of a gas is sufficiently high,
then Bose-Einstein condensation via sustainable evaporative cooling
is impossible. Changes in the collision rate and phase-space density
of the focused cloud are complicated by the fact that
alternate-gradient lensing automatically leads to an anisotropic
focused distribution both in space \textit{and} in velocity.

\begin{figure}[ht]
\begin{center}
\epsfxsize=\columnwidth \epsfbox{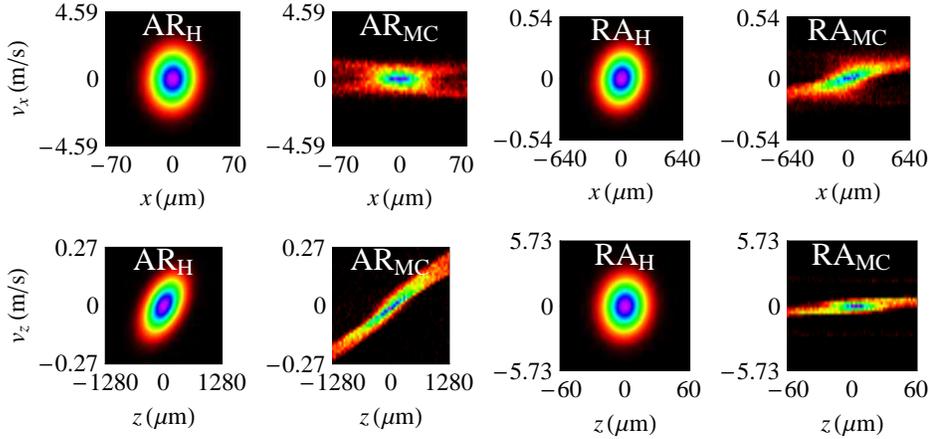}
 \caption{\label{phaseplot} Phase-space plots of the AR and RA
strategies from the rightmost points of Figures~\ref{altgrad} (a)
and (b) respectively. Both a harmonic lens calculation (subscript
$H$) and a Monte Carlo simulation with the full magnetic fields from
realistic coils (subscript MC) are plotted. In the AR$_{\rm MC}$
plot there are $45\%~(x-v_x)$ and $42\%~(z-v_z)$ of the initial
30,000 atoms present. In the RA$_{\rm MC}$ plot there are
$61\%~(x-v_x)$ and $24\%~(z-v_z)$ of the atoms present. The effects
of aberration are clearly seen when comparing the harmonic and
realistic magnetic coils.}
\end{center}
\end{figure}

At the $\mathcal{B}_{r,z}=0$ focus of the cloud the
$\mathcal{ABCD}_{r,z}$ matrix has radial and axial spatial widths
$\sigma_r=\langle r^2 \rangle^{1/2}=\mathcal{A}_{r}\sigma_R$ and
$\sigma_z=\langle z^2 \rangle^{1/2}=\mathcal{A}_{z}\sigma_R$
respectively. The radial and axial velocity widths are given by:
\begin{equation}
\sqrt{\langle {v_{r,z}}^2 \rangle}=\sqrt{k_B
\mathcal{T}_{r,z}/m}=\sqrt{{\mathcal{C}_{r,z}}^2
{\sigma_R}^2+{\sigma_V}^2/{\mathcal{A}_{r,z}}^2},
\end{equation}
where $\mathcal{T}_{r,z}$ is the atomic cloud temperature. In the
limit $\mathcal{C}_{r,z} {\sigma_R}\ll {\sigma_V}/\mathcal{A}_{r,z}$
(which is not always the case), the velocity width of the focused
cloud is inversely proportional to its spatial width and
$\sqrt{\langle {v_{r,z}}^2 \rangle}={\sigma_V}/\mathcal{A}_{r,z}$.
In Figure~\ref{phaseplot} phase-space plots of the AR and RA
strategies from the rightmost points of Figures~\ref{altgrad} (a)
and (b) were generated with a 30,000 atom simulation. The effect of
aberration is clearly seen when comparing the purely harmonic lenses
(subscript H) and the Monte Carlo simulation with the full magnetic
fields from realistic coils (subscript MC). The plots also
demonstrate the inversely proportional relationship between spatial
and velocity widths. Furthermore, in the Monte Carlo simulations
there is much stronger correlation between position and velocity.

The anisotropic temperature in the focused cloud means that captured
atoms in a cylindrically symmetric harmonic trap with radial and
axial frequencies $\nu_r$ and $\nu_z$, will rethermalise to an
isotropic temperature via elastic collisions. The total
(potential+kinetic) energy of the focused cloud is equated with that
of a 3D harmonic oscillator at equilibrium temperature $\mathcal{T}$
(i.e. $\frac{1}{2}k_B \mathcal{T}$ average energy per atom for each
space and velocity dimension). The equilibrium temperature of the
focused cloud after thermalisation is thus:
\begin{equation}
\mathcal{T}=\frac{m}{6k_B}\left(4\pi^2(2 {\nu_r}^2\langle
r^2\rangle+{\nu_z}^2\langle z^2\rangle)+(2\langle
{v_r}^2\rangle+\langle {v_z}^2\rangle)\right).
\end{equation}

In many experiments other physical properties of the atomic cloud
are of interest: the atomic density $n\propto
1/({\sigma_r}^2\sigma_z)$, collision rate $\gamma\propto n
\sqrt{\mathcal{T}}$ and phase-space density ${\rm PSD}\propto
n/(\mathcal{T}_r {\mathcal{T}_z}^{1/2})$. In order to minimise the
loss of phase-space density during thermalisation of the focused
cloud, one can show that it is best to choose $\nu_r$ and $\nu_z$
such that the potential energy is equal in all spatial dimensions
and the total potential energy is equal to the total kinetic energy
of the cloud.

The physical properties of the focused atomic clouds of
Figure~\ref{phaseplot} are displayed in Table~\ref{foctab}. The
relative values for the density $n,$ collision rate $\gamma$ and
phase-space density can be converted into absolute values using
typical initial experimental values $\langle n\rangle=10^{10}\,{\rm
cm}^{-3}$ (an atom number $N=3\times 10^7$ in the unfocused cloud),
$\gamma=1\,{\rm Hz}$ (for $^{87}$Rb with the low-temperature
collision cross-section $\sigma=8\pi a^2=8\times10^{-12}\,{\rm
cm}^2$) and phase-space density PSD=$2\times10^{-6}.$ Aberrations
will effectively be smaller if a large MOT is used with the same
density (e.g. $N=10^9,$ $\sigma_R=1.3\,$mm) and temperature as
Table~\ref{foctab}, and a trap with weaker frequencies to catch the
atoms.

\begin{table}[!th]
\begin{tabular}{|c|c|c|c|c|c|c|c|c|}
\hline
 &t&$P$&$\sigma_{r,z} (\mu$m)&$\mathcal{T}_{r,z} (\mu$K)&$\nu_{r,z} $(Hz)&$n$&$\gamma$&PSD\\
\hline
 &$0$&1&400, 400&20, 20&\textit{17.4, 17.4}&1&1&1\\
AR$_{\rm H}$&$T$&1&15.2, 257&13900, 60.9&\textit{12000, 47.3}&1080&23200&0.892\\
 &$T_R$&1&15.2, 257&9260, 9260&9850, 584&1080&23200&0.108\\
\hline
  &$0$&1.000&400, 400&20, 20&\textit{17.4, 17.4}&1&1&1\\
AR$_{\rm MC}$&$T$&0.308&23, 500&2720, 57&\textit{3530, 23.5}&74.5&713&0.325\\
 &$T_R$&0.308&23, 500&1830, 1830&2900, 133&74.5&713&0.085\\
\hline
&$0$&1&400, 400&20, 20&\textit{17.4, 17.4}&1&1&1\\
RA$_{\rm H}$&$T$&1&128, 12.2&197, 21600&\textit{170, 18700}&319&6110&0.988\\
 &$T_R$&1&128, 12.2&7320, 7320&1040, 10900&319&6110&0.046\\
\hline
 &$0$&1.000&400, 400&20, 20&\textit{17.4, 17.4}&1&1&1\\
RA$_{\rm MC}$&$T$&0.305&200, 20&105, 1670&\textit{79.7, 3180}&24.4&137&0.509\\
 &$T_R$&0.305&200, 20&627, 627&195, 1950&24.4&137&0.139\\
\hline

\end{tabular}
\caption{Physical properties of the two different alternate-gradient
strategies modelled, AR and RA from the rightmost points in
Figs.~\ref{altgrad} (a) and (b) respectively. Subscripts H and MC
respectively denote a simulation with purely harmonic lenses and a
Monte Carlo simulation with the full magnetic fields from realistic
lens coils. The measured parameters are: fraction of atoms in the
Gaussian focus $P$ (see text), radial/axial cloud radius
$\sigma_{r,z}$, radial/axial temperature $\mathcal{T}_{r,z}$,
radial/axial trap frequency $\nu_{r,z}$, relative density $n,$
relative collision rate $\gamma$ and relative phase space density
PSD. The effective trap frequencies for the initial $(t=0)$ and
focused $(t=T)$ cloud (italicised) are equilibrium values based on
$\nu=\frac{1}{2\pi\sigma}\sqrt{k_B \mathcal{T}/m}$. The actual
frequencies of the trap the atoms are loaded into at $t=T$ are
denoted in the $t=T_R$ lines (the cloud properties after $T$ plus
the thermalisation time).\label{foctab}}
\end{table}

Note that the Monte Carlo relative density increases in
Table~\ref{foctab} are lower than those in Figure~\ref{altgrad} by a
factor of $\approx 2$, but the fraction of atoms in the focus are
higher by a factor of $\approx 2$. In Figure~\ref{altgrad} the
density is estimated by measuring the fraction of atoms arriving at
the harmonic focus. Here, a 6-dimensional Gaussian phase-space fit
to the narrow central peak of the bimodal focus was made to
explicitly obtain $\sigma_r,$ $\sigma_z,$ $\mathcal{T}_r$ and
$\mathcal{T}_z.$ The fraction of atoms in this Gaussian focus was
used for $P$ here.

An interesting result of Table \ref{foctab} is that the aberrations
of `real' lenses work to our advantage, to some extent, in that the
atoms can be loaded into a trap with a shallower depth than atoms
focused by a purely harmonic lens, and phase-space density loss is
reduced during rethermalisation. This is due to the reduced
anisotropy of the spatial and velocity distributions at the focus.
The focused atoms have a relatively high temperature and one needs a
trap depth of $\approx 10\,$mK to trap the focused atoms. For an
atom with a magnetic moment of one Bohr magneton $\mu_B,$ this
corresponds to a magnetic trap depth $150\,$G.

\section{Discussion and conclusion}
\label{sec:5} The main application of interest for the magnetically
imaged atoms will be loading a magnetic microtrap or optical dipole
trap, for which alternate gradient imaging is well-suited. If the
trap that is being loaded is harmonic, with a large capture volume,
then the rms size of the cloud will be linked to the equilibrium
temperature after elastic collisions rethermalise the initially
bimodal image distribution. In order to keep the high density core
of the atomic cloud, the high energy atoms must be removed on a time
scale that is rapid compared to rethermalisation - this could be
achieved with strong RF evaporative cooling or by shining resonant
dark SPOT beams \cite{SPOT} at the focal region. A trap with a small
capture volume, e.g. an atom chip \cite{atomchips} or a focused
optical dipole beam trap \cite{dipole}, is ideal as only the high
density core of atoms will be captured in the trap.

The timescale needed to remove the shell of hot diffuse atoms from
the central core will actually be much longer than the time it takes
for the dense core atoms (which are anisotropically focused in space
and velocity) to thermalise to a uniform temperature in 3D. The
higher the anisotropy of the focused atomic cloud, the more phase
space density is lost during the thermalisation. This loss could be
mitigated to some extent if the radial and axial trap frequencies
were varied to remain in equilibrium with the changing radial and
axial cloud temperatures during the (short) thermalisation time.

In this paper we have used experimentally realistic parameters to
compare the limiting focal size of a launched cold cloud of
weak-field seeking atoms subject to either a single or double
magnetic lens impulse. The $\mathcal{ABCD}$ matrix formalism was
convenient for giving an estimate as to the parameters needed for
magnetic focusing, but numerical simulations were necessary to
detect the effects of aberrations in real magnetic lenses. If one
wishes to minimise the rms image volume of a launched cloud then a
single-impulse lens is preferable. If, however, one can selectively
capture the central core of the bi-modal image, a double-impulse
(alternate gradient) lens can lead to orders of magnitude relative
density increases for both pancake- and sausage-shaped image clouds.
Although we have only considered cold thermal atomic clouds in this
paper, the effects of aberrations will also be important for tight
focusing of coherent matter waves \cite{BECmirror,Patrik}.

\ack This work is supported by EPSRC, the UKCAN network and Durham
University. We thank Charles Adams, Simon Gardiner and Kevin
Weatherill for fruitful discussions.

\section*{Appendix: modelling time varying lens strengths}

This paper has only discussed parabolic lenses with constant
strength $\omega,$ pulsed on with a top-hat pulse of duration
$\tau$. In experiments the lens strength is a function of time,
partly because the current in a real coil is not a top-hat pulse,
and partly because the centre-of-mass of a launched atomic cloud
changes as it goes through a lens and thus (to second order in
position) will experience a time-varying parabolic lens. In practice
it was unnecessary to adjust the timing or lens coil positions to
allow for these effects in the simulations, however we briefly
discuss ways around this issue should it become problematic.

A harmonic lens with arbitrary time variation $\omega(t)$ (where
$\omega(t)$ and its derivatives are zero outside the experimental
pulse time $t_1\leq t\leq t_1+\tau_1$), is equivalent to an infinite
product of infinitesimal $2\times 2$ translation and thin lens
matrices, resulting in a single $2\times 2$ matrix that is itself
independent of the initial velocity and position of an atom. On
solving $y''[t]=-\omega^2[t] y[t]$ from $t=t_1$ to $t=t_1+\tau_1,$
with the initial conditions $\{y[t_1],y'[t_1]\}=\{dy,0\}$ to get
$\{y[t_1+\tau_1],y'[t_1+\tau_1]\}=\{\mathcal{A},\mathcal{C}\} dy$,
then use initial conditions $\{y[t_1],y'[t_1]\}=\{0,dv\}$ to get
$\{y[t_1+\tau_1],y'[t_1+\tau_1]\}=\{\mathcal{B},\mathcal{D}\} dv$,
(with small values for $dy$ and $dv$) results in the general
$\mathcal{A}\mathcal{B}\mathcal{C}\mathcal{D}$ matrix for any
initial position and velocity from (numerically or analytically)
solving the differential equation for only two different initial
conditions. As this $\mathcal{A}\mathcal{B}\mathcal{C}\mathcal{D}$
matrix has determinant 1 (it is a product of determinant 1 matrices)
it can be expressed as a (translation matrix)-(thin
lens)-(translation matrix) combination.

One can then use an iterative 4D Newton-Raphson method with four input parameters (the
$z$ positions of the two alternate gradient lenses and the times $t_1$ and $t_2)$ such
that the $z$ centre-of-mass velocity of the atoms is not altered by either lens, and
$\mathcal{B}_r$ and $\mathcal{B}_z$ are identical to zero.

\end{document}